\documentclass[aps,prb,showpacs, twocolumn, superscriptaddress]{revtex4}
\usepackage[utf8]{inputenc}
\usepackage[english]{babel}
\usepackage{graphicx}
\usepackage{color}
\usepackage{bm}
\usepackage{amssymb}
\usepackage[intlimits]{amsmath}
\usepackage{amsmath}

\begin{document}

\title{Casimir effect mechanism of pairing between fermions in the vicinity of a  magnetic quantum critical point}
\author{Yaroslav A. Kharkov}
\author{Oleg P. Sushkov}
\affiliation{School of Physics, University of New South Wales,   Sydney 2052, Australia}

\begin{abstract}
We consider two immobile spin $1/2$ fermions in a two-dimensional magnetic
system that is close to the $ O(3)$ magnetic quantum critical point (QCP)
which separates magnetically ordered and disordered phases.
Focusing on the disordered phase in the vicinity of the QCP, we demonstrate that the criticality results in a strong long range
attraction between the fermions, with potential
$V(r) \propto -1/r^{\nu}$, $\nu \approx 0.75$, where $r$ is separation between 
the fermions. The mechanism of the enhanced attraction is similar to Casimir effect and corresponds to multi-magnon exchange processes between the fermions.
While we consider a model system, the problem is originally motivated
by recent establishment of magnetic QCP in hole doped cuprates under the superconducting dome at doping of about 10\%.
We suggest the mechanism of magnetic critical enhancement of pairing in cuprates.
\end{abstract}
\maketitle

\section{Introduction}
In the present paper we study interaction between fermions mediated by magnetic fluctuations 
in a vicinity of magnetic quantum critical point. 
To address this generic problem we consider a specific model of two
holes injected into the bilayer antiferromagnet.
The results presented below demonstrate that critical magnetic fluctuations lead to the long range Coulomb-like attraction between the holes.

Our interest to this  problem is motivated by cuprates.
Lying at the center of the debate of high-$T_c$ superconductivity is whether it originates 
from a Fermi liquid or from a Mott insulator. Recent experimental data, including Angle-Resolved Photoemission Spectroscopy (ARPES) support Mott insulator scenario in underdoped cuprates and show transition from small to large Fermi surface in the hole doping range $0.1<x<0.15$, see Refs.~\cite{Hossain08,He11,Yang11} Magnetic quantum oscillations (MQO) in underdoped YBa$_2$Cu$_3$O$_{6+y}$ also support the small pocket scenario \cite{DoironLeyraud07}, in contrast to the large Fermi surface observed on the overdoped side \cite{Vignolle08}.
Besides that, existence of hole pockets is consistent with the picture of dilute holes 
dressed by spin fluctuations, based on doping a Mott insulator. \cite{Liu92}

Optimally doped and overdoped cuprates do not have any static magnetic order. On the other hand, the 
underdoped cuprates possess a static incommensurate magnetic order at zero temperature. 
A magnetic QCP separating these two regions was predicted in Ref.~\cite{Milstein08} at doping
$x\approx0.1$. 
In La$_{2-x}$Sr$_x$CuO$_4$ the QCP is  smeared out because of disorder. However, 
in YBa$_{2}$Cu$_{3}$O$_{6+y}$ the QCP is located experimentally with neutron scattering, nuclear magnetic resonance wipeout and muon spin rotation ($\mu SR$) at doping  $x \approx 0.09$ ($y\approx 0.47$). 
\cite{Stock08,Hinkov08,Haug10} At larger doping, after crossing the QCP the (quasi-) static magnetic ordering vanishes and becomes fully dynamic.

It is widely believed that superconducting pairing in cuprates is driven by a
magnetic mechanism. The most common approach is based on the spin-fermion 
model in the frame of normal Fermi liquid picture (large Fermi surface).
Within this approach electrons interact via exchange of an antiferromagnetic 
(AF) fluctuation (paramagnon).~\cite{Scalapino5}
The lightly doped AF Mott insulator approach, instead, necessarily implies small Fermi 
surface. In this case holes interact/pair via exchange of 
the Goldstone magnon.~\cite{kuchiev} Due to the strong on-site Hubbard
repulsion both approaches result in the d-wave pairing of fermions.

Magnetic criticality can significantly influence superconducting pairing.
This idea has been recently considered by Wang and Chubukov~\cite{wang} in a context of electron doped cuprates.
There are also some earlier works referenced in Ref.~\cite{krotkov}.
However, to the best of our knowledge all the previous works imply a normal
liquid with large Fermi surface. This might be a reconstructed Fermi surface
which emulates small hole pockets~\cite{moon}, but still in essence this is a weak
coupling normal Fermi liquid like approach. A large Fermi surface to a significant
extent diminishes importance of the magnetic criticality for the pairing.

In this work we consider two holes injected in the 2D ``rigid'' Mott insulator, so in 
essence our approach  implies the small Fermi surface and therefore strong coupling limit.
In this case influence of the magnetic criticality
on the coupling between two fermions is the most dramatic. 
As the Mott insulator host we use  the bilayer antiferromagnet with magnetic fluctuations 
driven by the interlayer coupling. 
We consider the bilayer model for the sake of
performing a controlled calculation. However, we believe
that conceptually our conclusions are equally applicable
to the single-layer and multi-layer cuprates.
The model presented here has only commensurate magnetic ordering,
so we put aside incommensurability in the cuprates.

The model under consideration demonstrates spin-charge separation at the QCP. \cite{Holt} It means delocalization of hole spin due to dressing by divergent magnon cloud. The effect of spin-charge separation points out to the nontriviality of the pairing problem. We are not familiar with any models that consider pairing of two fermions, that incorporate physics of spin-charge separation.

In order to probe the interaction between two fermions we consider spin fluctuations in the system, keeping the fermions to be immobile and spatially localized, just as magnetic impurities. \cite{Vojta00} Because of the immobility,  Fermi statistics of the impurities is not actually relevant for our results, however for shortness we will call the impurities as "fermions" hereafter.
Our calculations show that the single magnon exchange is getting irrelevant close to the QCP. Instead, we obtain strong inter-fermion attraction in singlet and triplet spin channel due to Casimir effect. \cite{Casimir48} Each of injected fermions (holes) builds up a "bag" of the quantum magnetic fluctuations. The fermions attract to each other, sharing common bag and reducing energy of the magnetic fluctuations inside of the bag. So-called "spin-bag" mechanism of attraction  in the context of high-$T_c$ superconductivity in cuprates is familiar from the works of J. Schrieffer \textit{et al.} \cite{Schrieffer88}, that however did not account for the magnetic criticality, neither small Fermi surface. The  spin bag model has also conceptual similarity to QCD bag models for nucleon binding such as MIT \cite{Chodos74} and chiral bags \cite{Brown79} that have being extensively studied from 1970's up to now.

The structure of the paper is following. In the Section \ref{sec:model} we introduce bilayer $J-J_\perp$ antiferromagnet, which is simple but instructive model and contains all essential physics of magnetic criticality. 
In the Section \ref{sec:Magnons_QCP} we characterize magnetic quantum critical point driven by interlayer coupling $J_\perp/J$ and describe magnon excitations for undoped AF in disordered phase in the frame of spin-bond mean field theory. Next, in the Section \ref{sec:Hole-magnon_interaction} we  move to the hole-doped $J-J_\perp$ model and show how holes interact with magnons. In the Section \ref{sec:Hole-hole_interaction_mediated_by magnons}, which is the main content of the paper, we consider hole-hole pairing problem at the QCP and show that pairing can not be described in terms of one-magnon exchange. In Section \ref{sec:Effective_theory_for Casimir-like_interaction} we develop effective theory for Casimir interaction of the fermions, considering double-fermion "atom" which can be either in singlet or triplet state. In Section \ref{sec:Solution_to_Dyson equation} we present results of solution to Dyson's equations for singlet and triplet Green's functions and finally show how binding energy in both spin channels depends on inter-fermion distance $r$. Finally, we draw our conclusions in the Section \ref{sec:Conclusions} and provide supplementary material in the Appendix.

%
\section{Model}\label{sec:model}

\begin{figure}[h]
\includegraphics[scale=0.5]{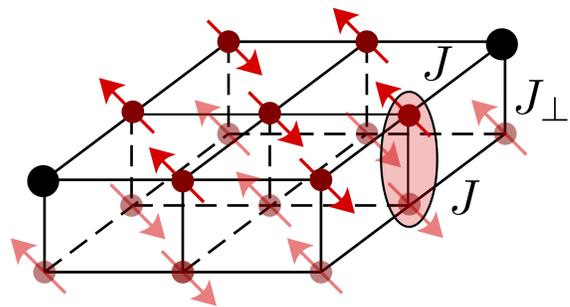}
\caption{Bilayer $J-J_\perp$ antiferromagnet model. Two black dotes on the top layer represent holes.\label{fig:bilayer_AF}}
\end{figure}

Our model system is $J-J_\perp$ square lattice bilayer Heisenberg antiferromagnet at zero temperature, where magnetic fluctuations are driven by interlayer coupling $J_\perp$ (see Fig. \ref{fig:bilayer_AF}).
The Hamiltonian of the undoped host AF reads 
\begin{eqnarray}
\label{eq:JJHamiltonian}
 H_{J,J_\perp}  &=&  J\sum_{\langle i,j\rangle}({\bf S}_i^{(1)}\cdot {\bf S}_j^{(1)} + {\bf S}_i^{(2)}\cdot {\bf S}_j^{(2)}) +\nonumber \\   && J_{\perp}\sum_{i}{\bf S}_i^{(1)}\cdot {\bf S}_i^{(2)}, 
\end{eqnarray}
\noindent 
The superscripts (1), (2) in Eq. \eqref{eq:JJHamiltonian} indicate the layers, $\langle i, j\rangle$ denotes summation over nearest neighbour sites. Here ${\bf S}_i^{(1)}=\frac{1}{2}c_{i\mu,1}^{\dagger}{\bm \sigma}_{\mu\nu}c_{i\nu,1}$ is spin of an electron at site $i$ on the top plane and $c_{i\sigma,1}^{\dagger}/c_{i\sigma,1}$ is creation/annihilation operator of an electron with spin  $\sigma=\uparrow,\downarrow$  at site $i$, $\bm \sigma_{\mu\nu}$ are Pauli matrices.  The Hamiltonian describes the antiferromagnetic coupling in the each layer as well as between the two layers. 
It is known that without holes (half-filling) the model has an $O(3)$ magnetic QCP at $J_{\perp}/J = 2.525$ (see Refs. \cite{Sandvik94, Sandvik95,Zheng97,Kotov98}) separating the AF ordered and the magnetically disordered phase of spin dimers. Note that since we consider zero temperature case, the magnetic ordering in the AF phase is consistent with the Mermin-Wagner theorem.  We dope the first layer with two holes. For simplicity we set hopping integrals equal to zero, therefore the holes are immobile. The holes interact with each other via magnetic fluctuations of the spins, i.e. exchanging by magnons. 

In the subsections A and B of the current section we will briefly present formalism, which describes magnon excitations and hole-magnon interaction on the basis of the bilayer model. For more detailed explanations see \cite{Holt}; a reader which is not interested in these technical details can go directly to the Section \ref{sec:Hole-hole_interaction_mediated_by magnons}.

\subsection{Magnons at QCP}\label{sec:Magnons_QCP}
Magnetic excitations in the magnetically disordered phase are magnons, which are also in literature called triplons. In the present paper we will use terms magnons and triplons as synonyms. To describe the magnons we employ the spin-bond operator mean field technique. This approach has been previously applied to quantum disordered systems such as bilayer antiferromagnets, spin chains, spin ladders,  Kondo insulators etc.\cite{Jurecka00, Eder98, Saito03, Vojta99, Matsushita99, Yu99}. It is known~\cite{Matsushita99, Yu99} that  this simple technique gives the position of the QCP at $(J_{\perp}/J)_{c} \approx 2.31$, which is close to the exact value $(J_{\perp}/J)_{c} = 2.525$ known from Quantum Monte Carlo calculations \cite{Sandvik94, Sandvik95}, series expansions \cite{Zheng97}, and involved analytical calculations with the use of the Brueckner technique \cite{Kotov98}. 
The spin-bond technique being much simpler than the Brueckner technique has sufficient accuracy for our purposes.

The bond-operator representation describes the system in a base of pairs of coupled spins on a rung, which can either be in a singlet or triplet (triplon) state. So, we define singlet $s^\dag_i$ and triplet $(t^\dag_{ix}, t^\dag_{iy}, t^\dag_{iz})$  operators that create  a state at site $i$ with spin zero and spin one, which is polarized along one of the axises $(x,y,z)$.
The four types of bosons obey the bosonic commutation relations. To restrict the physical states to either singlet or triplet, the above operators are subjected to the constraint

\begin{equation}
\label{eq:constraint}
s_i^\dagger s_i + \sum_{\alpha}t_{i\alpha}^\dagger t_{i\alpha} = 1.
\end{equation}

\noindent In terms of these bosons, the spin operators in each layer ${\bf S}_i^{(1)}$ and ${\bf S}_i^{(2)}$ can be expressed as 

\begin{equation}
\label{eq:spins}
S_{i\alpha}^{(1,2)} = \frac{1}{2}(\pm s_i^\dagger t_{i\alpha} \pm t_{i\alpha}^\dagger s_i - i\epsilon_{\alpha \beta \gamma}t_{i\beta}^\dagger t_{i\gamma}),
\end{equation} 
see Ref. \cite{ChubukovSachdev}.
Substituting the bond-operator representation of spins defined in Eq. \eqref{eq:spins} into the $H_{J,J_{\perp}}$ in Eq. \eqref{eq:JJHamiltonian} we obtain 

\begin{eqnarray}
\label{eq:bondoperatorHamiltonian}
&&H_{J,J_{\perp}} = H_1 + H_2 + H_3 + H_4,\nonumber\\
&&H_1 = J_{\perp}\sum_{i}\left(-\frac{3}{4}s_i^\dagger s_i + \frac{1}{4}t_{i\alpha}^\dagger t_{i\alpha}\right), 
\nonumber \\
&&H_2 = \frac{J}{2}\sum_{\langle i,j\rangle}(s_i^\dagger s_j^\dagger t_{i\alpha} t_{j\alpha} + s_i^\dagger s_j t_{i\alpha} t_{j\alpha }^\dagger + {h.c.}),\nonumber\\
&&H_3 = \frac{J}{2}\sum_{\langle i,j \rangle} i\epsilon_{\alpha \beta \gamma}( t_{j\alpha}^\dagger t_{i\beta}^\dagger t_{i\gamma} s_j +  {h.c.}),\nonumber\\
&&H_4 = \frac{J}{2}\sum_{\langle i,j \rangle}(t_{i\alpha}^{\dagger} t_{j\beta}^{\dagger} t_{i\beta} t_{j\alpha} -  t_{i\alpha}^{\dagger} t_{j\alpha}^{\dagger} t_{i\beta} t_{j\beta}).\nonumber\\
\end{eqnarray}
The Hamiltonian (\ref{eq:bondoperatorHamiltonian}) contains quadratic, cubic and quartic terms in magnon operators $t$. The most important for us are the quadratic terms, because they provide quantum criticality. The only effect due to the nonlinear terms $H_3$ and $H_4$ is renormalization of parameters near the QCP, such as position of the QCP, magnon velocity, magnon gap and etc. This does not affect physics at the QCP, and therefore we will neglect these terms in further considerations. 

The next step for treating the Hamiltonian (\ref{eq:bondoperatorHamiltonian}) is to account for the hard-core constraint (\ref{eq:constraint}). It could be done by introducing infinite on-site repulsion of triplons; however, this technique is quite involved.  Another, more simple, way is to employ mean-field approach, accounting for the constraint (\ref{eq:constraint}) via a Lagrange multiplier $\mu$ in the Hamiltonian
\begin{equation}
H_{J,J_\perp} \rightarrow H_{J,J_\perp} - \mu \sum_i(s_i^\dagger s_i + t_{i\alpha}^\dagger t_{i\alpha} - 1).
\end{equation} 
Further analysis is straightforward. We replace singlet operators by numbers, $\langle s_{i}^{\dagger} \rangle = \langle s_{i} \rangle = \bar{s}$ (Bose-Einstein condensation of spin singlets); and diagonalize the quadratic in $t$ Hamiltonian by performing the usual Fourier and Bogoliubov transformations

\begin{equation}
\label{eq:Fourier}
t_{i\alpha} = \sqrt{\frac{1}{N}}\sum_{\bm q}e^{i \bm{q r_{i}}}  \left( u_{q} b_{q\alpha} + v_{q} b_{-q\alpha}^{\dagger}\right).
\end{equation}

\noindent Here $N$ is the number of spin dimers in the lattice; the diagonalized Hamiltonian reads
\begin{eqnarray}
\label{eq:diagonalizedmfa}
H_{m}(\mu, \bar{s}) &=& E_0(\mu, \bar s)
+ \sum_{\bm q}\omega_{q}b_{q\alpha}^\dagger b_{q\alpha},
\end{eqnarray}
\noindent where $\omega_{q} = \sqrt{A_q^2 - 4 B_q^2}$ and coefficients $A_q = \frac{J_{\perp}}{4} - \mu + 2J\bar{s}^2\gamma_q$, $B_q = J\bar{s}^2 \gamma_q$. Here we define
\begin{equation}
\gamma_q = \frac{1}{2}\left(\cos(q_x)+\cos(q_y)\right).
\end{equation} 
The lattice spacing is set to unity. 
The ground state energy  
\begin{equation}
E_0(\mu, \bar{s}) = N\bigg (-\frac{3J_{\perp} s^2}{4} - \mu \bar{s}^2 + \mu \bigg ) + \frac{3}{2}\sum_{\bm q}(\omega_{q} - A_{q})
\end{equation}
just shifts energy scale, and therefore is irrelevant for our purposes. 
The Bogoliubov coefficients $u_q$ and $v_q$  are given by
\begin{eqnarray}
\label{eq:bogoliubovcoeff}
u_q = \sqrt{\frac{A_q}{2\omega_{q}} + \frac{1}{2}}, \quad
v_q = -\mathrm{sign}(B_{q})\sqrt{\frac{A_q}{2\omega_{q}} - \frac{1}{2}}.
\end{eqnarray}
\noindent The parameters $\mu$ and $\bar{s}$ are determined by the saddle point equations:
$\partial E_0(\mu, \bar s)/\partial \mu = \partial E_0(\mu, \bar s)/\partial \bar{s} = 0$.
Solution to these equations gives position of the QCP at $J_\perp/J = 2.31$ and values of "chemical potential" $\mu=-2.706$ and singlet density $\bar s = 0.906$. We see that even at the QCP $\bar s$ is close to unity, which again justifies smallness of the nonlinear terms $H_3$ and $H_4$ in the Hamiltonian.

The dispersion of magnons is 
\begin{equation}
\omega_k = \sqrt{ c^2({\bf k - Q})^2 + \Delta^2},\quad \bf Q = (\pi, \pi),
\end{equation}
in the vicinity of  the wave-vector ${\bm Q}$, 
here  $\Delta$ is the magnon gap and $c$ is the velocity of magnons $c = 2J\bar{s}^2 = 1.64 J$, where the more precise value  is $c=1.9 J$, see Ref. \cite{Zheng97}. In the AF ordered phase the magnons are Goldstone bosons and thus necessarily gapless. On the contrary, in the disordered phase the gap opens up and the spin-bond approach gives $\Delta\propto (J_\perp-J_{\perp,c})$ which is not far from the prediction for ${O}(3)$ universality class systems $\Delta\propto (J_\perp-J_{\perp,c})^\eta$ with critical index $\eta = 0.71$ (see Ref. \cite{Zinn-Justin}). So, the spin-bond method provides a reasonably accurate description of the  QCP.

\subsection{Hole-magnon interaction}\label{sec:Hole-magnon_interaction}
We dope our system with two immobile holes, by removing two electrons from the upper plane of the bilayer antiferromagnet. Hence we define the hole creation operator $a_{i\sigma}^{\dagger}$ with spin projection $\sigma = \uparrow,\downarrow$ by its action on the spin singlet bond $|s\rangle$

\begin{eqnarray}
\label{eq:holon}
a_{i\uparrow}^{\dagger}| s\rangle = c_{i\uparrow,2}^{\dag}| 0 \rangle, \qquad
a_{i\downarrow}^{\dagger} | s\rangle = 
c_{i\downarrow,2}^{\dag}| 0 \rangle \ ,
\end{eqnarray} 
\noindent where $| 0 \rangle$ is vacuum. 
The electron creation/annihilation operator in the upper plane can be expressed in terms of hole creation/annihilation operators  $a_{i\sigma}^{\dagger}/a_{i\sigma}$ (see Ref. \cite{Jurecka00}), and after substitution in (\ref{eq:JJHamiltonian}) it gives following part of the Hamiltonian which describes hole-magnon interaction
\begin{eqnarray}\label{eq:hole_Hamiltonian_full}
H_{hm} &=& - \frac{J\bar s}{2}\sum\limits_{\langle i,j\rangle}  \left \{ ({\bf t}_{j} +
{\bf t}^\dag_{j})\boldsymbol{\sigma}_{i} + ({\bf t}_{i} +
{\bf t}^\dag_{i})\boldsymbol{\sigma}_{j}\right\} - \nonumber\\ &&\frac{J}{2} \sum\limits_{\langle i,j\rangle} i ({\boldsymbol \sigma_i [{\bf t}_j^\dag\times {\bf t}_j}] + \boldsymbol{\sigma}_j [{\bf t}_i^\dag\times {\bf t}_i]).
\end{eqnarray}
\noindent Here ${\boldsymbol \sigma}_{i} = a^\dag_{i \mu} \boldsymbol{\sigma}_{\mu\nu} a_{i\nu}$.
The first line in the Hamiltonian (\ref{eq:hole_Hamiltonian_full}) corresponds to hole-magnon interaction vertex. The terms, describing hole-double-magnon vertex,  which come from the second line of (\ref{eq:hole_Hamiltonian_full})  will be neglected below, because they are irrelevant in the infrared limit. Performing again standard Fourier and Bogoliubov  transformations (\ref{eq:Fourier}) the Hamiltonian  (\ref{eq:hole_Hamiltonian_full}) can be rewritten as 
\begin{eqnarray}\label{eq:Hole-magnon}
H_{hm} \approx \sum_i \sigma_{i}^\alpha\sum_{\bm q}  g_q(b_{q\alpha} e^{i{\bf qr}_i} + b^\dag _{q\alpha}e^{-i{\bf qr}_i}).
\end{eqnarray}
The hole-magnon vertex is equal to 
\begin{equation}
g_q = -\frac{J\bar s}{\sqrt{N}} \gamma_q(u_q + v_q).
\end{equation}
Note, that at the QCP the vertex diverges at ${\bf q}\rightarrow {\bf Q}=(\pi,\pi)$, because of the singularity of Bogoliubov coefficients $u_q,v_q \propto 1/\sqrt{\omega_q}\rightarrow\infty$.

The divergence of $g_q$ is crucial for the physics of fermion-magnon coupling at the QCP. In fact, it results in the phenomenon of spin-charge separation for the single-fermion problem. \cite{Holt} The spin of the hole is distributed in the power-law divergent cloud of magnons and in this sense is separated from the charge of the hole, localized on the hole's site. Later in the paper we will show that the infrared divergence of the fermion-magnon coupling at the QCP results in strong power-law attraction between the fermions. 

\begin{figure}
\includegraphics[width=8cm]{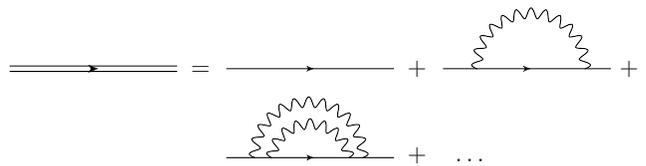}
\begin{picture}(0,0)
  \put(-162,33){\text{$=$}}
  \put(-80,33){\text{$+$}}
  \put(0,33){\text{$+$}}
  \put(-80,0){\text{$+ \quad \ldots$}}    
\end{picture}
\caption{Dyson's equation for single hole Green's function in Self-Consistent Born Approximation. Solid and waivy lines correspond to hole and magnon Green's functions.}
\label{fig:G_equation}
\end{figure}

Importance of spin-charge separation for single-fermion problem at the QCP could be seen from analysis of analytical structure of the hole Green's function. 
Standard approach in order to calculate one-fermion Green's function is to use $1/\mathcal{N}$ expansion for the ${O}(\mathcal{N})$ group, where $\mathcal{N}=3$ is the number of magnon components. Summation of leading terms in the expansion arises in Self-Consistent Born Approximation (SCBA), see Fig. \ref{fig:G_equation}.

\begin{figure}[h]
\includegraphics[scale=0.3]{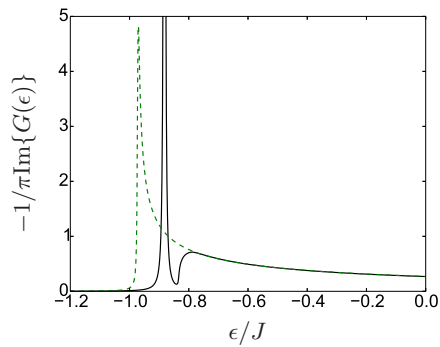}
\begin{picture}(0,0)
  \put(-180,35){ \rotatebox{90}{\text{$-1/\pi \mathrm{Im}\{G(\epsilon)\}$}}}
  \put(-95,-5){\text{$\epsilon/J$}}
\end{picture}
\caption{ Spectral function  $-1/\pi \, \mathrm{Im}\{G(\epsilon)\}$ of a single immobile hole obtained in SCBA (see \cite{Sushkov00}). The  green dashed curve corresponds to the QCP ($\Delta = 0$), and the black solid  line corresponds to the magnon gap $\Delta = 0.1 J$. Note that at the QCP the quasiparticle pole disappears.}
\label{fig:Sushkov_G_function}
\end{figure}

Calculations of the hole Green's function in the disordered phase in SCBA have been performed in Refs. \cite{Vojta00, Sushkov00}.
The results show that away from the QCP, in the disordered magnetic phase, quasiparticle pole in the fermion Green's function is separated by $\Delta$ from incoherent part of the Green's function. But when approaching to the QCP the Green's function instead of normal pole has just branch cut singularity. This is a consequence of the infrared singularity of the coupling constant $g_q$.

Spectral density of the fermion Green's function (see Fig. \ref{fig:Sushkov_G_function}) has inverse square root behaviour $G(\epsilon)\propto 1/\sqrt{\epsilon_0-\epsilon}$ in the vicinity of the singularity point $\epsilon_0$ and quasiparticle residue is approaching to zero $Z\propto \sqrt{\Delta}$ at the QCP. Here 
\begin{equation}
\epsilon_0\approx -0.97 J\label{eq:epsilon_0}
\end{equation}
is the position of the branching point of the Green's function and has a meaning of fermion energy shift due to interaction with magnons, where we set bare energy of the hole to zero.


\section{Hole-hole interaction, mediated by magnons}\label{sec:Hole-hole_interaction_mediated_by magnons}
Now we are ready to move to the actual problem of magnon mediated pairing of fermions and demonstrate new results. 
Adding up magnon Hamiltonian $H_m$, Eq. (\ref{eq:diagonalizedmfa}) and hole-magnon interaction Hamiltonian $H_{hm}$, Eq. (\ref{eq:Hole-magnon}), we arrive to effective Hamiltonian for two interacting holes, located at the sites with coordinates $\bm r_1$ and $\bm r_2$,
\begin{eqnarray}
\label{eq:H_eff}
H_{eff} &=& \sum_{\bm q} \omega_q b^\dag_{q \alpha} b_{q \alpha} + \nonumber\\ &&\sum_{i=1,2}\sigma^{\alpha}_{i} \sum_{\bm q} g_q (b_{q\alpha} e^{i\bm{qr}_i} + b^\dag_{q\alpha} e^{-i\bm{qr}_i}).
\end{eqnarray}
The Hamiltonian (\ref{eq:H_eff}) is applicable only if distance between holes $r=|\bm{r_1} - \bm{r_2}|>1$, as long as we put aside direct exchange interaction between two neighbouring holes. 

The effective model (\ref{eq:H_eff}) can be formulated in the language of a field theory. In fact, it is equivalent to the problem of two spin $1/2$ fermions coupled to a vector field $\bm{\phi}(\bm{r})$, described by $O(3)$-symmetric theory with Lagrangian
\begin{eqnarray}\label{eq:Lagrangian_phi4}
\begin{split}
\mathcal{ L } =& \frac{1}{2} (\partial_t \bm{\phi})^2  - \frac{c^2}{2} (\bm\nabla \boldsymbol{\phi})^2 - \frac{\Delta^2}{2} \bm{\phi}^2 -\\
 &\lambda (\bm{\phi}(\bm r_1)\bm\sigma_{1} + \bm{\phi}(\bm r_2)\bm\sigma_{2}),
 \end{split}
\end{eqnarray}
where  $\lambda$ is the coupling constant of fermion spin to magnon field. We focus only on the disordered magnetic phase, and therefore assume $\Delta^2\geq 0$. Let us note that in Eq. (\ref{eq:H_eff}) we neglected self-action of magnons, hence term $\propto\bm \phi^4$ is dropped in Eq. (\ref{eq:Lagrangian_phi4}).

Parameters of the Lagrangian (\ref{eq:Lagrangian_phi4}) could be directly expressed via parameters of the initial lattice Hamiltonian (\ref{eq:JJHamiltonian}). As an example, the coupling constant $\lambda$ in (\ref{eq:Lagrangian_phi4}) is related to the hole-magnon vertex  $g_q$ in the effective Hamiltonian (\ref{eq:H_eff}) as $g_q = \lambda/\sqrt{2\omega_q}$. Hence, for the Heisenberg bilayer model $\lambda\approx 2J\sqrt{c}$. This equivalence shows, that the problem of fermion pairing at the QCP, we are considering here, is generic. It has implications  far beyond the particular bilayer model.

\begin{figure}[h]
\includegraphics[scale=0.5]{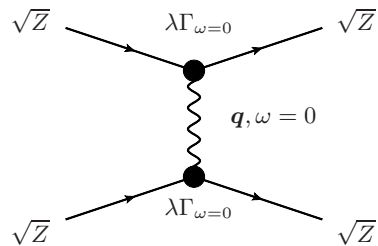}
\begin{picture}(0,0)
  \put(-72,85){\text{$\lambda\Gamma_{\omega=0}$}}
  \put(-72,15){\text{$\lambda\Gamma_{\omega=0}$}}
  \put(-47,50){\text{$\bm q, {\omega=0}$}}
  \put(-7,85){\text{$\sqrt{Z}$}}
  \put(-7,5){\text{$\sqrt{Z}$}} 
  \put(-130,5){\text{$\sqrt{Z}$}}
  \put(-130,85){\text{$\sqrt{Z}$}}            
\end{picture}
\caption{One-magnon exchange diagram that provides fermion-fermion interaction potential $V^{(1)}_{int}(\bm q)$.   Note that renormalization factor $\sqrt{Z}$ should be referred to each fermion line. Fermion-magnon vertices also come renormalized $\lambda\rightarrow \lambda\Gamma_{\omega=0}$. \label{fig:Born_diagr}}
\end{figure}

\begin{figure}[h]
\includegraphics[scale=0.4]{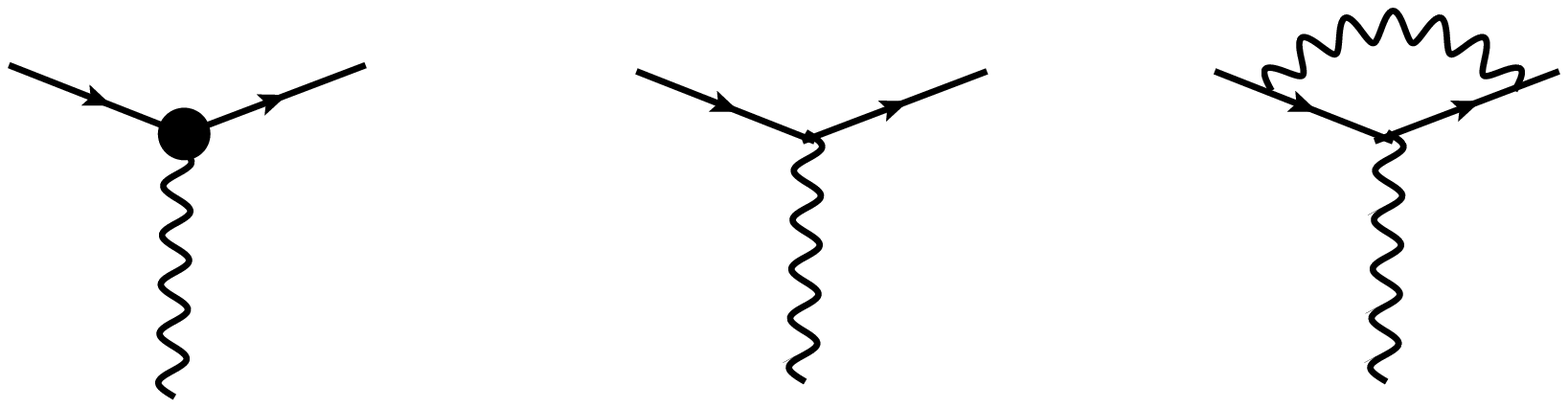}
\begin{picture}(0,0)
  \put(-176,40){\text{$\lambda\Gamma_\omega$}}          
  \put(-160,0){\text{$\bm q, \omega$}}            
  \put(-135,20){\text{$=$}}          
  \put(-65,20){\text{$+$}}            
  \put(5,20){\text{$+\, \ldots$}}              
\end{picture}
\caption{The fermion-magnon vertex.\label{fig:Vertex}}
\end{figure} 

\begin{figure}[h]
\includegraphics[width=0.35\columnwidth]{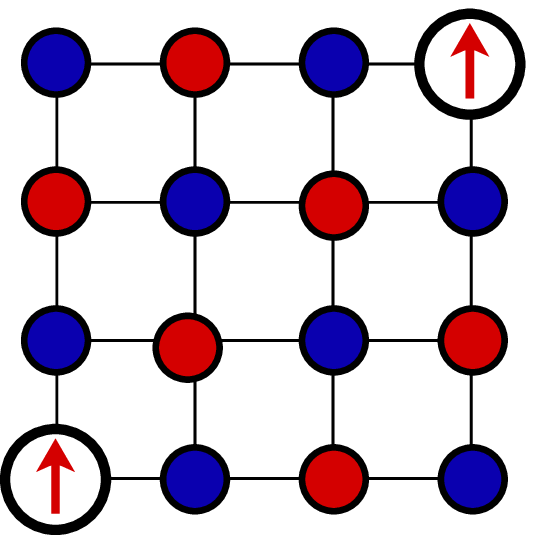}\hspace*{1cm}
\includegraphics[width=0.35\columnwidth]{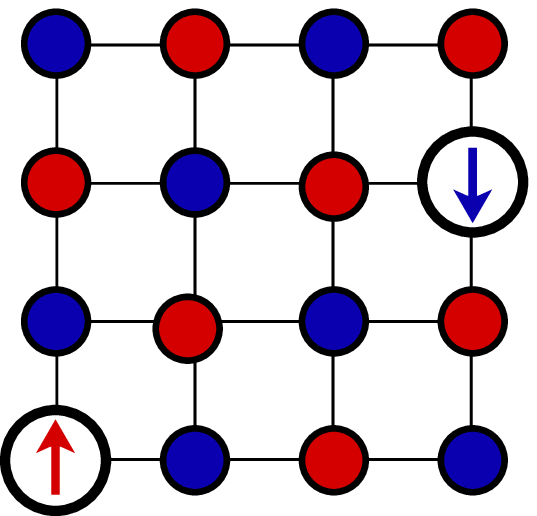}
\caption{Dependence of spin channel which provides attraction between holes on a mutual positioning of the holes in the lattice. Two holes with spins up symbolicaly represent triplet channel which provides negative interaction energy for  $P_r = (-1)^{r_x+r_y}= +1$, two holes with opposite spins represent singlet channel which results in attraction when $P_r = -1$ .\label{fig:fermion_favor_spin}}
\end{figure}

In order to calculate pairing energy between two fermions we first consider one magnon exchange contribution, Fig. \ref{fig:Born_diagr}. According to Feynman rules we obtain interaction potential 
\begin{equation}
V^{(1)}_{int}(\bm q) = - \lambda^2 Z^2\Gamma_{\omega=0}^2 \frac{\langle\bm\sigma_1 \bm\sigma_2\rangle}{c^2({\bf q}-{\bf Q})^2 +\Delta^2}.
\end{equation}
The factor $Z^2$ comes from $Z^{1/2}$ for each external fermion line. The vertex $\lambda \Gamma_{\omega=0}$ comes from diagrams in Fig. \ref{fig:Vertex}, $\lambda\rightarrow \lambda\Gamma_{\omega=0}$. Here $\omega$ is the frequency of the exchange magnon, which is equal to zero.
In the coordinate representation the potential reads
\begin{equation}\label{eq:V_Born_r}
V^{(1)}_{int}(r) = -\frac{\lambda^2}{2\pi c^2} Z^2\Gamma_{\omega=0}^2\cos(\bm{Q r})  \langle\bm\sigma_1 \bm\sigma_2\rangle K_0\left(\frac{r\Delta}{c}\right),
\end{equation}
where $K_0$ is the Macdonald function of zero's order. The potential energy $V^{(1)}_{int}(r) \propto \ln(r)$ is logarithmic at small distances $r<c/\Delta$ and  it exponentially decays at $r>c/\Delta$ as $V^{(1)}_{int}(r) \propto e^{-r\Delta/c}$. Spin-dependent prefactor $\langle\bm\sigma_1 \bm\sigma_2\rangle = 2[S(S+1) - 3/2]$ is determined by the total spin of two fermions $S$ and equals to  $-3$ in singlet channel and $+1$ in triplet channel. The potential is attractive in the state with total spin zero (one) when 
\begin{equation}
P_r=\cos({\bm{ Q r}})=(-1)^{r_x+r_y}
\end{equation}
is negative (positive) and repulsive in the opposite case (${\bf r} = r_x{\bf e_x} + r_y{\bf e_y}$).  This fact has clear physical meaning and reflects AF character of spin correlations in the antiferromagnet. The system tends to restore AF ordering and the state when the spins  of two interacting holes are aligned according to antiferromagnetic pattern (see Fig. \ref{fig:fermion_favor_spin}) is energetically preferable.

When we approach the QCP the quasiparticle residue as well as the magnon-hole vertex tends to zero: $Z\propto\sqrt{\Delta}\rightarrow0$ and $\Gamma_{\omega=0}\propto\Delta^{1/6}\rightarrow 0$ (see discussion in Section \ref{sec:Hole-magnon_interaction} and  Ref. \cite{Sushkov00}). Thus the single magnon exchange contribution given by (\ref{eq:V_Born_r}) vanishes, because the potential $V^{(1)}_{int}$ is proportional to $Z^2\Gamma_{\omega=0}^2\rightarrow 0$. 
Does this imply that pairing between fermions becomes very weak close to the QCP? Our answer is "no", on the contrary the pairing becomes very strong, but it is due the Casimir effect mechanism.

An approach to evaluate Casimir interaction energy between two spins can be drawn from analogy with calculation of Van der Vaals force between two atoms. The most elegant way to do so was first developed by Dzyaloshinsky. The Van der Vaals potential is given by "box diagram" with two-photon exchange between the atoms. \cite{Dzyaloshinsky56}

\begin{figure}[h]
\includegraphics[scale=0.5]{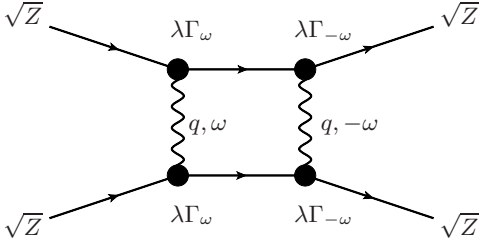}\\
\begin{picture}(0,0)
  \put(-27,80){\text{$\lambda\Gamma_{\omega}$}}
  \put(-27,10){\text{$\lambda\Gamma_{\omega}$}}
  \put(20,80){\text{$\lambda\Gamma_{-\omega}$}}
  \put(20,10){\text{$\lambda\Gamma_{-\omega}$}}  
  \put(-20,45){\text{$q, {\omega}$}}
  \put(30,45){\text{$q, {-\omega}$}}  
  \put(75,85){\text{$\sqrt{Z}$}}
  \put(75,5){\text{$\sqrt{Z}$}} 
  \put(-90,5){\text{$\sqrt{Z}$}}
  \put(-90,85){\text{$\sqrt{Z}$}}            
\end{picture}
\caption{The box diagram for two-magnon exchange between fermions. Note that the renormalization factor $\sqrt{Z}$ should be referred to each external fermion line. \label{fig:Box_diagr}}
\end{figure} 

We can try to follow this approach to calculate Casimir pairing energy between two fermions in the vicinity of the QCP. The diagram is shown in the Fig. \ref{fig:Box_diagr}. However, like in one magnon exchange diagram, we have to account for renormalization of external fermion lines, that results in additional factor $Z^2\rightarrow 0$. In such logic we again obtain zero for pairing energy at the QCP, and therefore we need another technique to solve the problem.
\subsection{The "Lamb shift" technique for calculation of Casimir interaction}\label{sec:Effective_theory_for Casimir-like_interaction}
In this section we introduce a new technique to treat Casimir pairing energy.
To incorporate "Casimir effect" physics, we consider composite two-fermion "atom", which has total spin either zero (singlet state) or one (triplet state). Next, we calculate "Lamb shift" in energy of this composite "atom" due to radiation of magnons as a function of separation between fermions.

Let's consider effective theory for the composite object. Creation operator for singlet state is 
\begin{equation}\label{eq:Psi_singlet}
\Psi^\dag_S = \frac{1}{\sqrt 2}(a^\dag_{1\uparrow}a^\dag_{2\downarrow} - a^\dag_{1\downarrow}a^\dag_{2\uparrow})
\end{equation}
and for triplet state
\begin{equation}\label{eq:Psi_triplet}
\begin{split}
&\Psi^\dag_{T,x} = \frac{-1}{\sqrt 2}(a^\dag_{1\uparrow}a^\dag_{2\uparrow} - a^\dag_{1\downarrow}a^\dag_{2\downarrow}),\\
&\Psi^\dag_{T,y} = \frac{i}{\sqrt 2}(a^\dag_{1\uparrow}a^\dag_{2\uparrow} + a^\dag_{1\downarrow}a^\dag_{2\downarrow}),\\
&\Psi^\dag_{T,z} = \frac{1}{\sqrt 2}(a^\dag_{1\uparrow}a^\dag_{2\downarrow} + a^\dag_{1\downarrow}a^\dag_{2\uparrow}).
\end{split}
\end{equation}
According to the selection rules  for interaction of the "atom" with magnon, there are three types of transitions $S\rightarrow T_\alpha$, $T_\alpha\rightarrow T_\beta$ and $T_\alpha\rightarrow S$, where $S$ means singlet state and $T_{\alpha}$ denotes triplet state with polarization $\alpha$.
The only one invariant kinematic structure that provides coupling between $S$ and $T, \alpha$ states with emission (absorption) of one magnon is $ \left\{ g_{ST}(q) \Psi^\dag_{T,\alpha}\Psi_S(b_{q\alpha} + b^\dag_{q\alpha}) +  h.c. \right\}$. In similar way, transition of the type $T_\alpha\rightarrow T_\beta$ is governed by the term $i g_{TT}(q)\varepsilon_{\alpha\beta\gamma}\Psi^\dag_{T,\alpha}\Psi_{T,\beta}(b_{q\gamma} + b^\dag_{q\gamma})$.  The coefficients $g_{ST}(q)$ and $g_{TT}(q)$ are coupling constants for these transitions. Therefore, the interaction of two-fermion system with magnon field in the singlet-triplet representation reads
\begin{eqnarray}\label{eq:Hamiltonian_singlet-triplet}
\mathcal{H} &=& \left\{ \delta_{\alpha\beta}\Psi^\dag_{T,\alpha}\Psi_S \sum_{\bm q} g_{ST}(q) (b_{q\beta} + b^\dag_{q\beta})+  h.c.\right\} + \nonumber\\
&&i\varepsilon_{\alpha\beta\gamma}\Psi^\dag_{T,\alpha}\Psi_{T,\beta} \sum_{\bm q} g_{TT}(q) (b_{q\gamma} + b^\dag_{q\gamma}).
\end{eqnarray}
The effective vertices can be calculated by evaluating matrix elements of the Hamiltonian (\ref{eq:H_eff}) between states (\ref{eq:Psi_singlet}), (\ref{eq:Psi_triplet}) :
\begin{eqnarray}
\begin{split}
g_{ST}(q) &= g_{TS}^*(q) = 2i g_q\sin \left(\frac{\bm{ q r}}{2}\right),\\ g_{TT}(q) &= 2g_q\cos\left(\frac{\bm{q r}}{2}\right).
\end{split}
\end{eqnarray}

\begin{figure}
\hspace*{-2.5cm}\includegraphics[width=0.35\columnwidth]{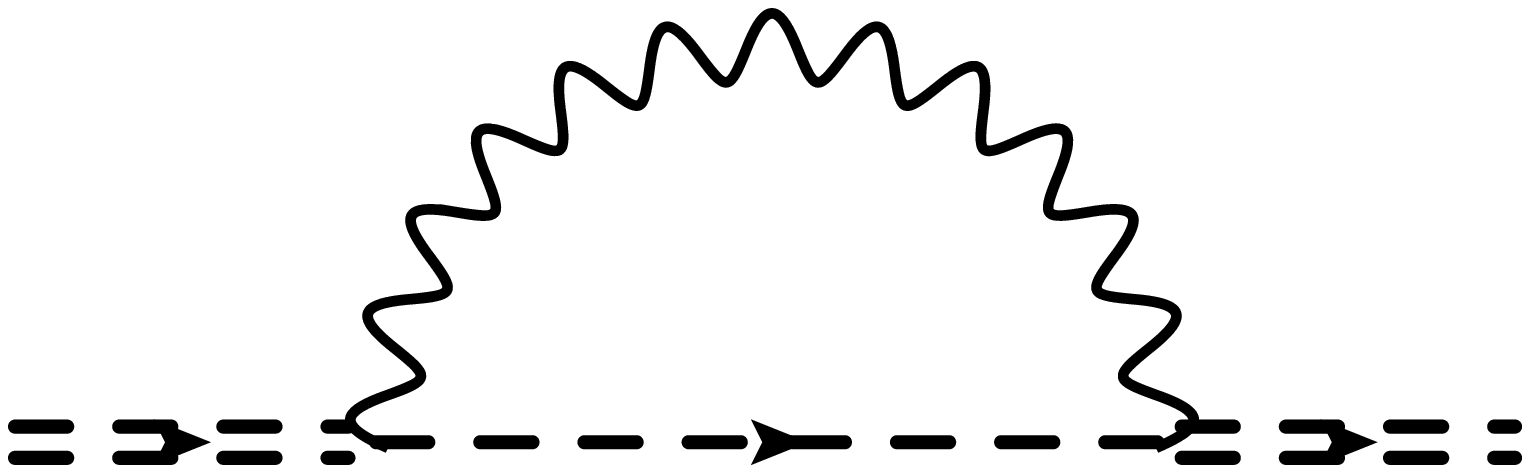}\\ \vspace*{0.5cm}
\hspace*{1.5cm}\includegraphics[width=0.8\columnwidth]{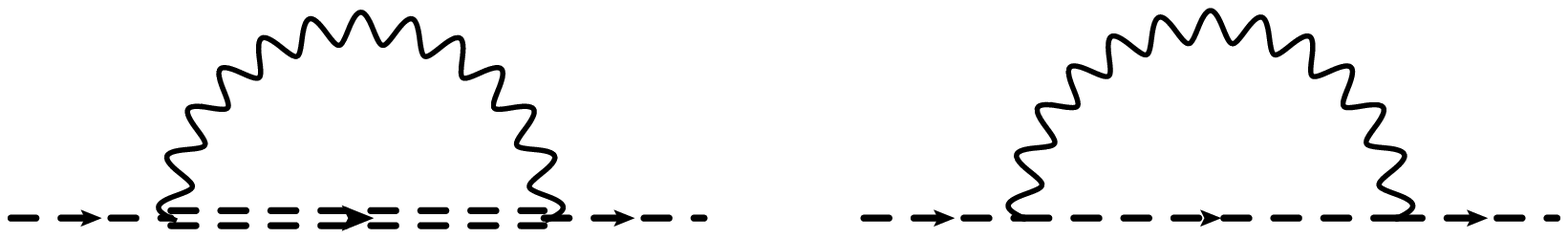}
\begin{picture}(0,0)
  \put(-245,50){\text{$\Sigma_S (\epsilon) = $}}
  \put(-245,5){\text{$\Sigma_T (\epsilon) = $}}  
  \put(-105,5){\text{$+$}}    
\end{picture}
\caption{Diagrams for singlet  $\Sigma_S(\epsilon)$ and triplet  $\Sigma_T(\epsilon)$ self-energies in SCBA.  Double dashed line and dashed line represent double-fermion Green's function in singlet and triplet channels correspondingly. Analytical expressions for the diagrams are given in Eq. (\ref{eq:Bag_Physics_Dyson}). \label{fig:Sigma_SCBA}}
\end{figure}

Let us define retarded Green's function for the singlet and triplet state
\begin{eqnarray}\label{eq:Green's_ST_funct_def}
&&G_{T, \alpha\beta}(t_2 - t_1) = -i \langle 0| \Psi_{T,\beta}(t_2) \Psi^\dag_{T,\alpha}(t_1)| 0\rangle \theta(t_2-t_1),\nonumber\\
&& G_{S}(t_2 - t_1) = -i \langle 0| \Psi_S(t_2) \Psi^\dag_S(t_1)| 0\rangle \theta(t_2-t_1),
\end{eqnarray}
where $|0\rangle$ is a ground state of the system and theta-function is
\begin{equation}
\theta(t) = \left\{
\begin{split}
&1,\, t>0;\\
&0,\, t<0.
\end{split} \right.
\end{equation}
Due to the $O(3)$ rotational invariance the triplet Green's function should be of the form $G_{T, \alpha\beta}(t) = \delta_{\alpha\beta} G_T(t)$.
Note that our definition of the Green's functions 
$G_{S}(t_2-t_1)$, $G_{T}(t_2 - t_1)$ assumes that the fermions, which constitute the composite "atom", are both created at the same moment of time $t_1$ and then both annihilated at the moment $t_2$. 
Fourier transform of Eq. (\ref{eq:Green's_ST_funct_def}) gives the Green's functions in the frequency representation
\begin{equation}
\begin{split}
&G_{S,T}(\epsilon) =\int_0^\infty dt e^{i(\epsilon + i\eta)t}G_{S,T}(t),
\end{split}
\end{equation}
where $\eta = +0$. Dyson's equations for singlet and triplet state Green's functions read
\begin{equation}\label{eq:G_01_Dyson}
G_{S,T}(\epsilon) = \frac{1}{\epsilon - \Sigma_{S,T}(\epsilon)+i\eta}.
\end{equation}
We use SCBA to evaluate singlet and triplet self-energy
\begin{equation}\label{eq:Bag_Physics_Dyson}
\begin{split}
\Sigma_S(\epsilon) =& 3\sum_{\bm q} |g_{ST}(q)|^2  G_T(\epsilon - \omega_q), \\
\Sigma_T(\epsilon) =& \sum_{\bm q} |g_{ST}(q)|^2 G_S(\epsilon - \omega_q) + \\ &2\sum_{\bm q}   |g_{TT}(q)|^2 G_T(\epsilon - \omega_q).
\end{split}
\end{equation}
The diagrams for the singlet and triplet self-energies are presented in Fig. \ref{fig:Sigma_SCBA}.
The combinator factors here come from contraction of the corresponding tensor structures of the coupling vertices in (\ref{eq:Hamiltonian_singlet-triplet}) and have a meaning of the number of the polarizations of intermediate state.

\subsection{Solution to Dyson's equations for singlet and triplet states and hole-hole interaction energy}\label{sec:Solution_to_Dyson equation}
In order to find interaction energy of two fermions, we numerically solve the system of two Dyson's equations (\ref{eq:Bag_Physics_Dyson}) in the square Brillouin zone for different inter-fermion separations $r$, measured in units of lattice spacings.  Energy grid in our computation is $\Delta\epsilon = 10^{-3} J$.
The zero approximation Green's function is 
$G^{(0)}_{S,T}(\epsilon) = 1/(\epsilon + i\eta)$
and for artificial broadening we take $\eta = 10^{-3}J$.
In order to perform numerical integration in equations (\ref{eq:Bag_Physics_Dyson}) we can directly integrate over square Brillouin zone, or introduce effective momentum cutoff  and integrate analytically over angle in momentum  space and then integrate over radial component of the momentum $|\bm q'| = |\bm q - \bm Q|\leq\Lambda_q \approx 1$ numerically. We have checked that there is a good agreement  between these two methods. However, the effective momentum cutoff method is much more efficient for numerics and provides better precision of the computations, therefore we mostly used the later approach.

The limit $r\rightarrow\infty$ of infinite separation between the fermions  corresponds to the case, when the vertices   in the equations (\ref{eq:Bag_Physics_Dyson}) are subsituted  to the averaged ones over $q$ oscillations $|g_{ST}(q)|^2, |g_{TT}(q)|^2 \rightarrow 2 |g_q|^2$ . The position of singularity of triplet and singlet Green's functions gives us energy $E_\infty $ of the two-fermion system separated by infinite distance. It is clear that in such limit the Green's functions in both spin channels should coincide $G_S(\epsilon) = G_T(\epsilon)$ which guarantees the same value for the asymptotic energy $E_\infty$ in singlet and triplet states. So, we refer interaction energy as a difference $V_{int}(r)=E(r)-E_\infty $.

\begin{figure}[h]
\begin{center}
{\includegraphics[width=0.85\columnwidth]{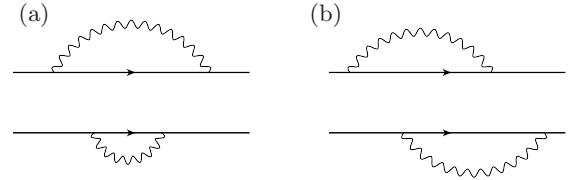}}
\begin{picture}(0,0)
  \put(-210,60){\text{(a)}}
  \put(-100,60){\text{(b)}}
\end{picture}
\caption{Diagrams contributing to $E_\infty$. Top and bottom solid line  correspond to single hole Green's function. Diagram (a) is accounted in SCBA (\ref{eq:Bag_Physics_Dyson}), diagram (b) is not included in (\ref{eq:Bag_Physics_Dyson}) and corresponds to $1/\mathcal{N}$ correction to SCBA.}\label{fig:E_infty}
\end{center}
\end{figure}

We found that the value $E_\infty \approx -1.55 J$ (at the QPC)  is about $20 \%$ smaller compared to doubled energy of an isolated single hole $2\epsilon_0 \approx -1.94 J$, see Eq. (\ref{eq:epsilon_0}). This difference is due to the fact that certain diagrams, which are presented in SCBA for single hole Green's function, are not included in SCBA for the two hole Green's function $G_{S,T}$ (see Fig. \ref{fig:E_infty}). This deviation shows precision of our method, which can be improved by calculating $1/\mathcal{N}$ corrections to SCBA (\ref{eq:Bag_Physics_Dyson}).


As it's seen from the structure of effective vertices
\begin{equation}
\begin{split}
&|g_{ST}(q)|^2 = 2 g_q^2 (1 - P_r\cos{\bm{q'r}}),\\
&|g_{TT}(q)|^2 = 2 g_q^2(1 + P_r \cos {{\bm{q' r}}}),
\end{split}
\end{equation}
the system's behaviour greatly depends on the "parity" $P_r=(-1)^{r_x+r_y}$ of the inter-fermion distance. The holes prefer to form singlet (triplet) spin state for negative (positive) "parity" $P_r$ at given $r$.

\begin{figure}[h]
\begin{center}
{\includegraphics[width=0.49\columnwidth]{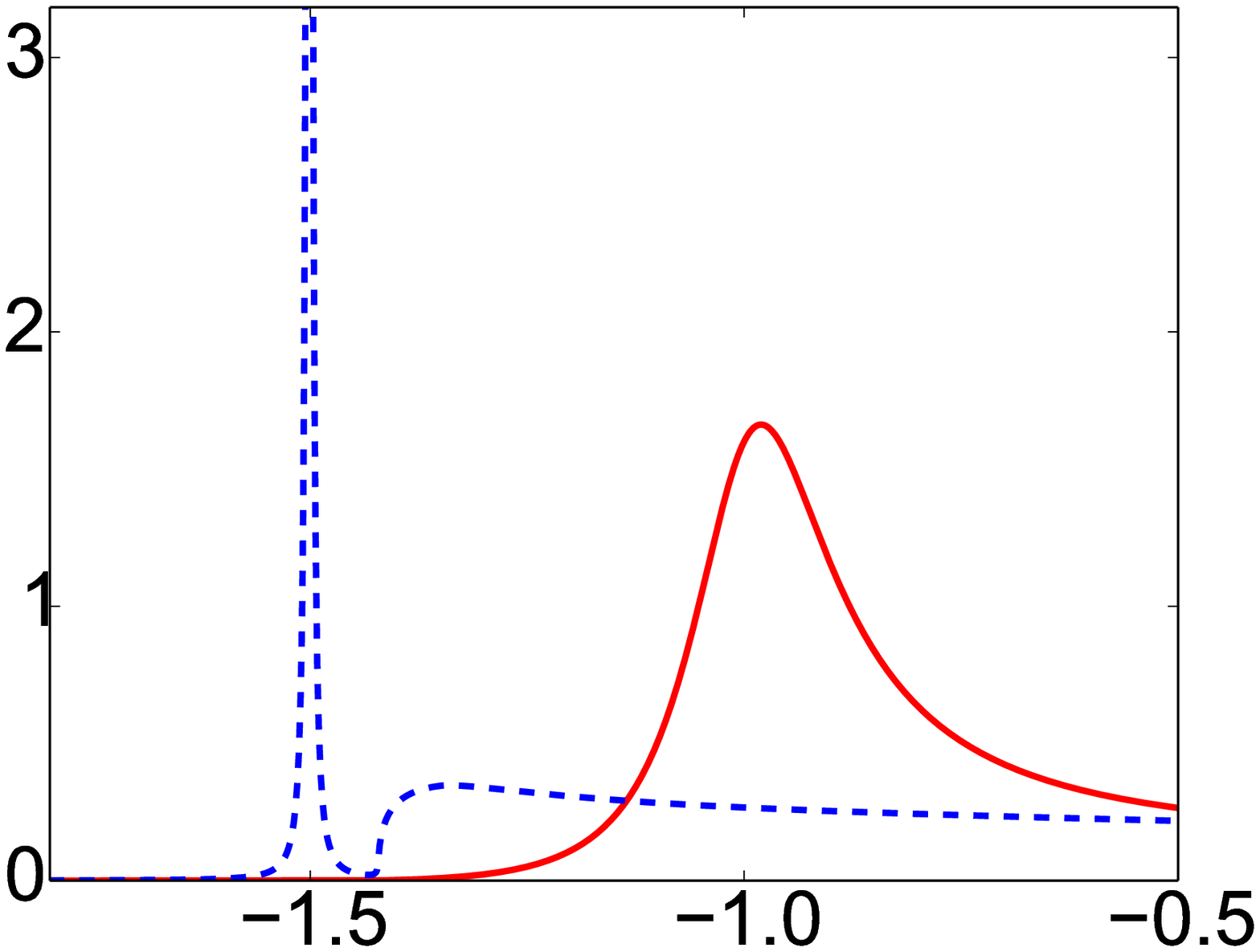}}
{\includegraphics[width=0.49\columnwidth]{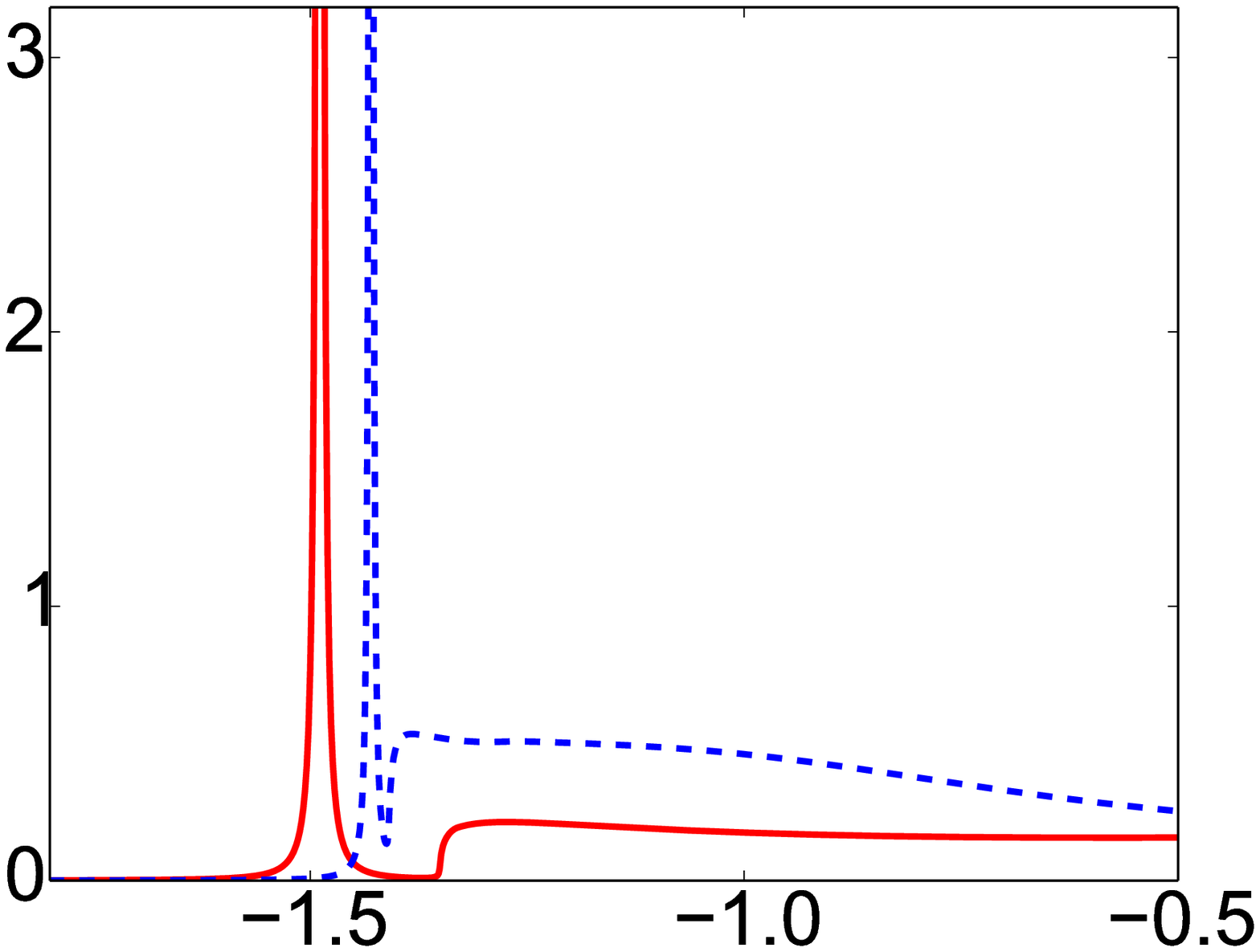}}
\begin{picture}(0,0)
  \put(-130,40){ \rotatebox{90}{\text{$A_{S,T}(\epsilon)$}}}
  \put(-60,0){\text{$\epsilon/J$}}
  \put(-5,40){ \rotatebox{90}{\text{$A_{S,T}(\epsilon)$}}}
  \put(60,0){\text{$\epsilon/J$}}
  \put(-30,80){\text{(a)}}
  \put(95,80){\text{(b)}}
  \put(-60,80){\text{$r=4$}}
  \put(65,80){\text{$r=5$}}  
\end{picture}
\caption{ Spectral functions $A_S(\epsilon)$, $A_T(\epsilon)$ of double-fermion Green's functions in singlet and triplet channels close to the QCP ($\Delta = 0.08J$).
Panel (a) corresponds to inter-hole distance $r=4$ and (b) corresponds to $r=5$. Blue dashed lines correspond to triplet state, red solid lines correspond to singlet state.\label{fig:Green's_funcs_Finite_Gap}}
\end{center}
\end{figure}

First, let us consider the case, when the system is away from the QCP, $\Delta>0$. 
We plot spectral functions
\begin{equation}
A_{S,T}(\epsilon)=-\frac{1}{\pi}\, \mathrm{Im} \{ G_{S,T}(\epsilon) \}  
\end{equation}
for singlet and triplet Green's functions at different $r$, see Fig.\ref{fig:Green's_funcs_Finite_Gap} (a, b). 
We see well defined quasiparticle peak in the triplet (singlet) spin channel at $r=4$ ($r=5$). However, in the opposite spin channel the peak broadens and submerges to continuum. This effect can be interpreted as a formation of an excited decaying state, which coupled via magnons to the ground  state.


%

\begin{figure}
\begin{center}
\includegraphics[width=0.99\columnwidth]{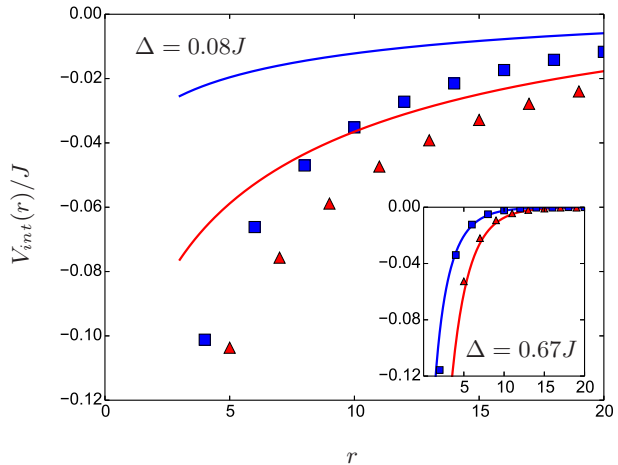}
\end{center}
\begin{picture}(0,0)
  \put(-130,90){ \rotatebox{90}{\text{$V_{int}(r)/J$}}}
  \put(0,15){\text{$r$}} 
  \put(-80,170){\text{$\Delta=0.08J$}}
  \put(45,55){\text{$\Delta=0.67J$}}   
\end{picture}
\caption{Interaction energy of two holes $V_{int}(r)$  at finite magnon gaps as a function of inter-hole distance $r$. Red trangles and blue squares show the results of the "Lamb shift" technique in singlet and triplet state. Red and blue solid lines represent theoretical prediction from one-magnon exchange mechanism (\ref{eq:V_Born_r}) for singlet and triplet channels. The main plot corresponds to  small magnon gap ($\Delta = 0.08 J$), the inset corresponds to large magnon gap ($\Delta=0.67 J$).\label{fig:Binding_energy_Finite_Gap}}
\end{figure}

In Fig. \ref{fig:Binding_energy_Finite_Gap} we plot the fermion-fermion interaction energy versus distance. The inset displays the interaction energy when the system is away from the QCP (the magnon gap is large, $\Delta = 0.67 J$). Squares and triangles show results of our "Lamb-shift" technique calculations and solid lines represent the single magnon exchange formula (\ref{eq:V_Born_r}). There is an excellent agreement between the two approaches. The main part of Fig.  \ref{fig:Binding_energy_Finite_Gap} shows the same quantities, but close to the QCP (the magnon gap is small, $\Delta = 0.08 J$). Here we observe a dramatic disagreement between the result of the "Lamb-shift"  technique and the single magnon exchange potential (\ref{eq:V_Born_r}). The single magnon exchange approximation fails in the vicinity of the QCP.

%

\begin{figure}
\begin{center}
{\includegraphics[width=0.65\columnwidth]{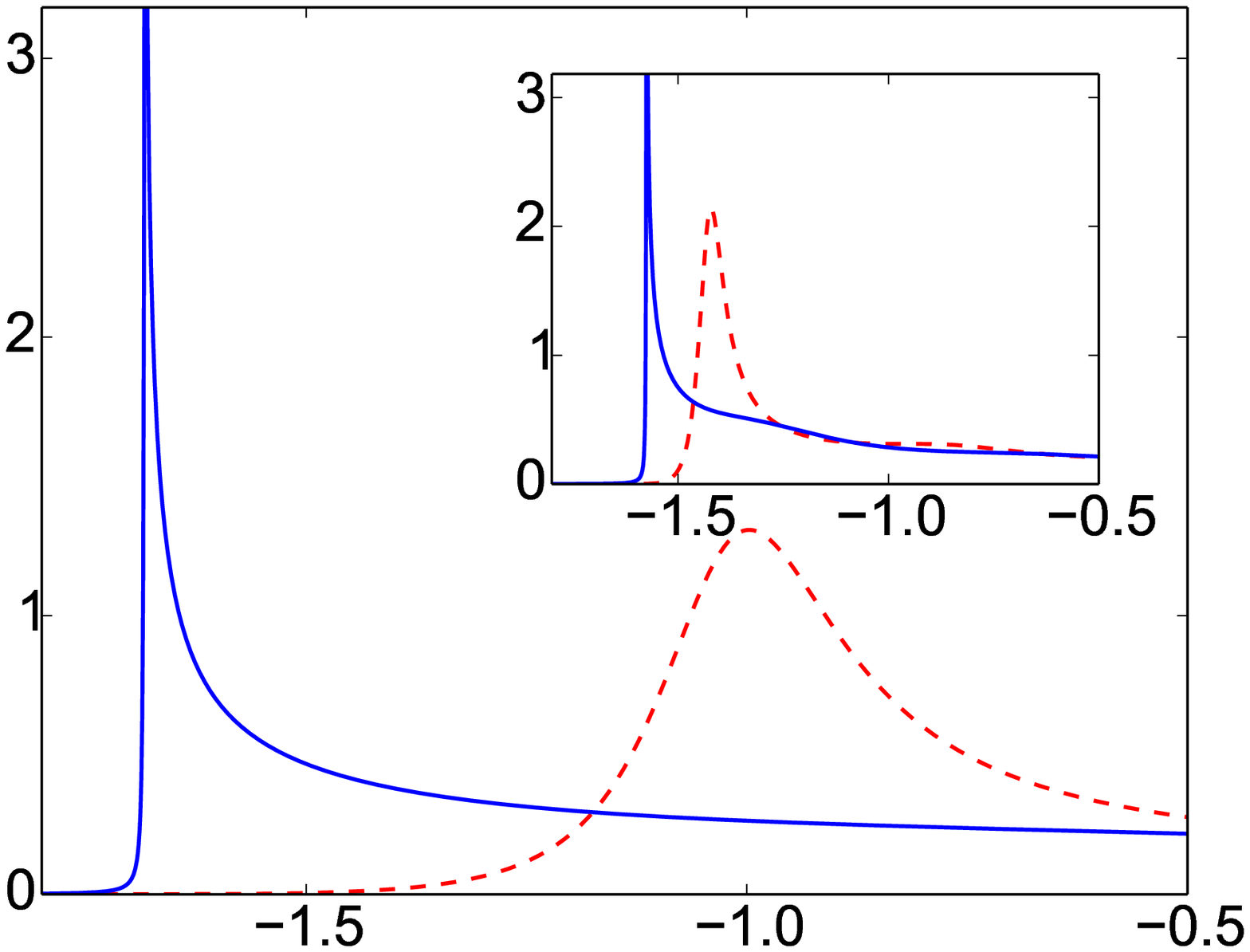}}
{\includegraphics[width=0.65\columnwidth]{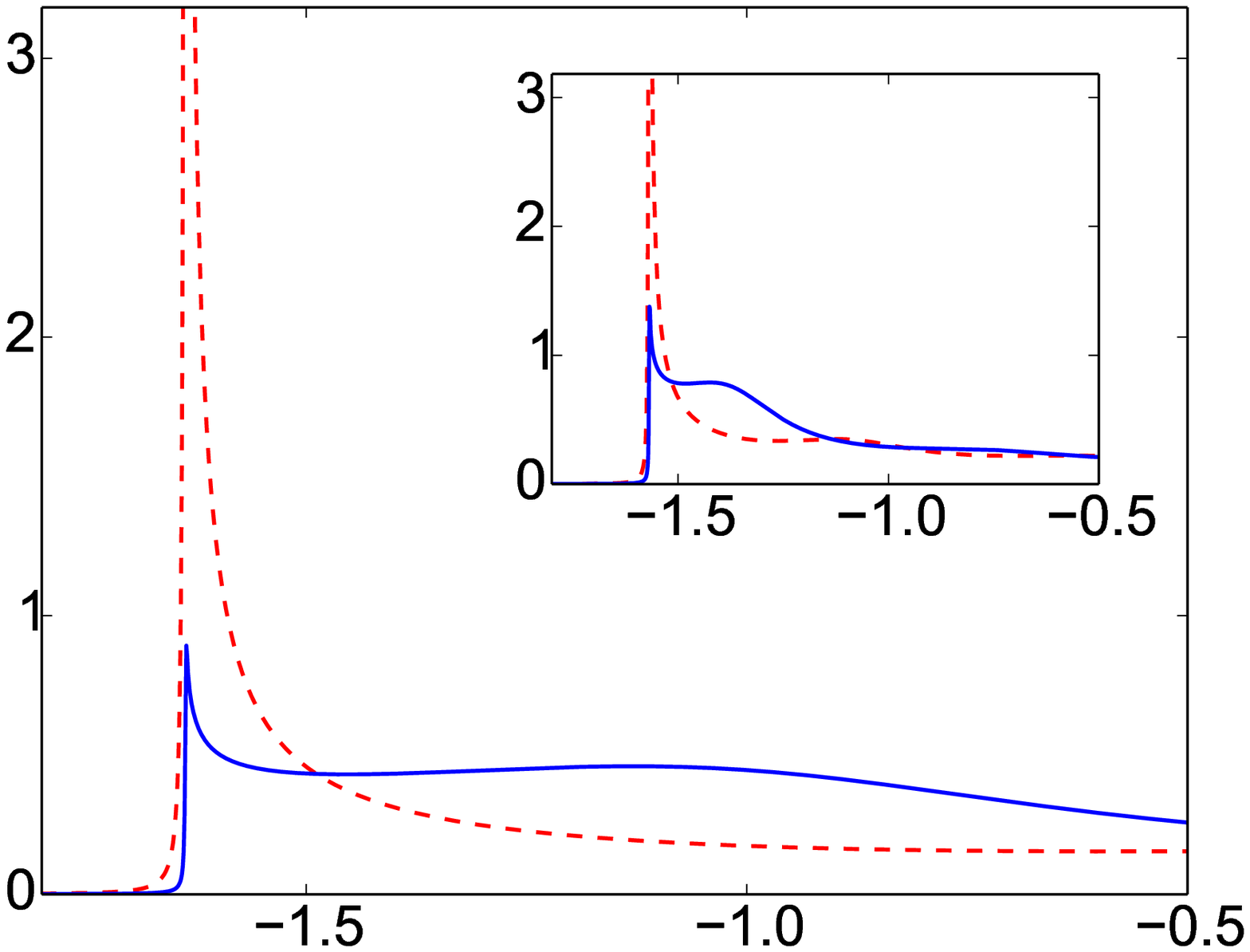}}
\begin{picture}(0,0)
  \put(-170,40){ \rotatebox{90}{\text{$A_{S,T}(\epsilon)$}}}
  \put(-90,0){\text{$\epsilon/J$}}
  \put(-170,160){ \rotatebox{90}{\text{$A_{S,T}(\epsilon)$}}}
  \put(-90,120){\text{$\epsilon/J$}}
  \put(-120,215){\text{(a)}}
  \put(-120,93){\text{(b)}}
  \put(-120,200){\text{$r=4$}}
  \put(-120,77){\text{$r=5$}} 
  \put(-55,210){{\text{$r=20$}}}
  \put(-55,87){{\text{$r=21$}}} 
\end{picture}
\caption{Spectral functions $A_S(\epsilon)$, $A_T(\epsilon)$ of double-fermion Green's functions in singlet and triplet channels at the QCP ($\Delta=0$). On the panel (a) the main plot corresponds to inter-hole distance $r=4$, the inset plot to $r=20$.  On the panel (b) the main plot corresponds to $r=5$, the inset plot corresponds to $r=21$. Red solid lines show  singlet state, blue dashed lines show triplet state.\label{fig:Green's_funcs}}
\end{center}
\end{figure}

Let us now consider the most interesting case of pairing between fermions at the QCP ($\Delta = 0$). As in the case of a single fermion at the QCP, the Green's functions $G_S(\epsilon)$ and $G_T(\epsilon)$  have just power-law cuts, instead of quasiparticle peaks, with branching point $E=E(r)$, see Fig. \ref{fig:Green's_funcs} (a, b).
The position $E(r)$ of the branching point gives the ground state energy of the system. Spin channel of the ground state is specified by the spin state in which the Green's function is singular at $E(r)$. The state, in which the Green's function is not singular, corresponds to decaying state.
Imaginary part of both singlet and triplet Green's functions emerges at the same branching point $E(r)$ for any fixed $r$ (see Fig. \ref{fig:Green's_funcs}). This is due to 
transitions between the states with emission of soft magnons with $\omega_q \rightarrow 0$. In the Fig. \ref{fig:Green's_funcs} (b) we see distinct discontinuity of both singlet and triplet spectral functions at the same branching point $E(r)$. In the Fig. \ref{fig:Green's_funcs} (a) the branching points are also coincide, but singlet spectral function has small spectral weight in the vicinity of the branching point.


\begin{figure}[h]
\begin{center}
\includegraphics[width=0.99\columnwidth]{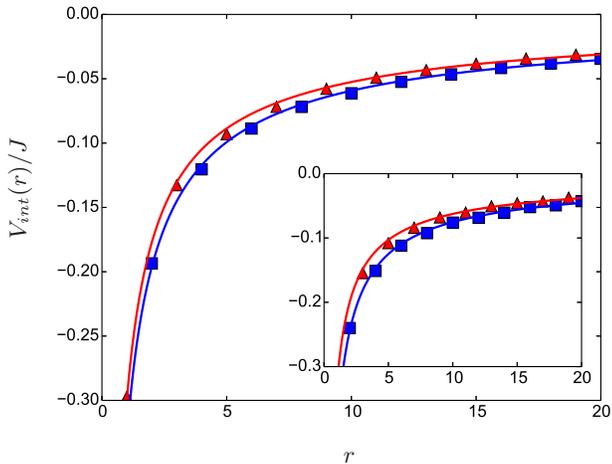}
\end{center}
\begin{picture}(0,0)
  \put(-130,100){ \rotatebox{90}{\text{$V_{int}(r)/J$}}}
  \put(0,15){\text{$r$}}    
\end{picture}
\caption{Interaction energy $V_{int}(r)$ of two holes at the QCP, $\Delta=0$. 
Red trangles and blue squares show the results of the "Lamb shift" technique for singlet and triplet state.
Red and blue solid lines represent power-law fits $V_{int}(r) = - a/r^\nu$ for singlet and triplet channel. The main plot corresponds to SCBA, in the inset we represent $V_{int}(r)$, which includes first  $1/\mathcal{N}$  correction to the SCBA. The exponent for all curves is approximately $\nu\approx0.75$. \label{fig:Binding_energy}}
\end{figure}

Our results for the interaction energy $V_{int}(r)$ at the QCP  as a function of distance $r$, obtained within the "Lamb shift" technique, are presented in Fig. \ref{fig:Binding_energy}. 
We see from the data that the interaction between two fermions is attractive, when the "parity" $P_r$ is negative (positive), see Fig. \ref{fig:fermion_favor_spin}. The binding becomes stronger at smaller inter-fermion distances $r$. The interaction energy has a power-law form 
\begin{equation}
V_{int}(r)=-a/r^\nu, \quad \nu\approx0.75
\end{equation}
with prefactor $a\approx0.3 J$, where $\nu$ and $a$ are found from the least-square fit of our numerical data. The values for prefactor $a$ and power exponent $\nu$ are slightly different for singlet ($a=0.3 J$, $\nu = 0.76$) and triplet ($a=0.33 J$, $\nu = 0.74$) cases. The variations of the values of $a$ and $\nu$ are negligible within the accuracy of our calculations.

In the inset in Fig. \ref{fig:Binding_energy} we show $V_{int}(r)$ which includes vertex corrections to SCBA (\ref{eq:Bag_Physics_Dyson}). Leading in $1/\mathcal{N}$ corrections for the  singlet and triplet self-energy $\delta\Sigma_S$ and $\delta\Sigma_T$ are presented in Appendix (see diagrams in Fig. \ref{fig:delta_Sigma} and formulas (\ref{eq:delta_Sigma_singlet}), (\ref{eq:delta_Sigma_triplet})). 
These corrections increase binding by about $20\%$, leaving critical index $\nu$ almost unchanged. 
Thus we conclude that corrections in $1/\mathcal{N}$ to SCBA do not change qualitative and quantitative picture, given by SCBA. 


From our calculations we observe very strong long-range attraction between fermions in the vicinity of the QCP. We clearly see that one magnon exchange contribution to the interaction energy vanishes at the QCP. On the contrary, accounting for multi-magnon exchange processes we obtain significant binding in singlet and triplet channels. We calculate the attraction energy due to multi-magnon exchange processes as a "Lamb shift" of energy of a two fermion "atom" due to emission of multiple magnons. The fermions interact, sharing common "bag" of magnetic fluctuations and reducing energy of fluctuations inside of the "bag". Therefore, the physics of inter-fermion attraction in the vicinity of the QCP is due to "Casimir bag" mechanism.

\section{Conclusions}\label{sec:Conclusions}
In conclusion, we considered interaction between two spin $1/2$ fermions embedded in a two dimensional antiferromagnetic system at the QCP, which separates ordered and disordered magnetic phases. As a model system we study bilayer antiferromagnet at $T=0$ with two injected holes, in which magnetic criticality is driven by interlayer coupling. We have shown that in the vicinity of the QCP the interaction between fermions can not be described by simple one-magnon exchange, unlike the case when the system is away from the QCP. The interaction mechanism is similar to Casimir effect and is due to multi-magnon exchange processes. To incorporate features of Casimir physics we developed a new approach, which we call a "Lamb shift" technique. We considered composite two-fermion "atom" and calculated it's energy shift ("Lamb shift") provided by radiation of magnons. We found strong attraction between the fermions in spin singlet and triplet states depending on the  "parity" of the inter-fermion distance $r$, which is positive (negative) for even (odd) $r$. Positive (negative) "parity" corresponds to attraction in triplet (singlet) channel. The attractive potential has  power-law form $V(r)\propto -1/r^\nu$ with the exponent $\nu\approx0.75$. 

We suppose that our work sheds light on the influence of magnetic criticality on fermion pairing mediated by magnons. We also believe that our results are conceptually applicable to cuprates.

\section{Aknowledgements}
We gratefully acknowledge A. Chubukov, G. Khaliullin and I. Terekhov for useful discussions. This research was supported by Australian Research Council (Grant No. DP110102123).

\newpage
\appendix
\section{Leading $1/\mathcal{N}$ corrections to SCBA for two-fermion Green's function}

\begin{figure}[h]
\hspace*{1.7cm}\includegraphics[width=0.75\columnwidth]{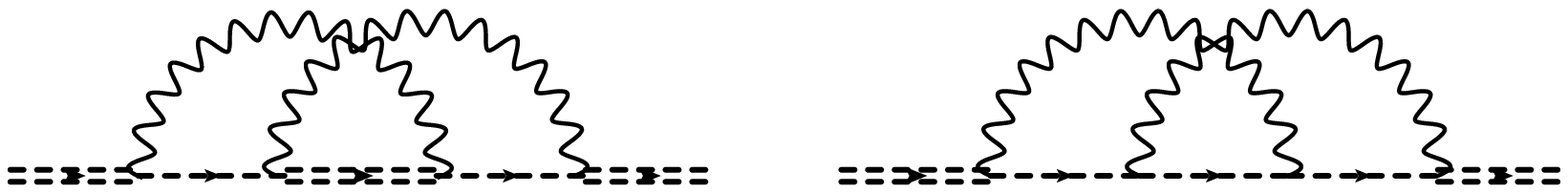}
\hspace*{1.7cm}\includegraphics[width=0.75\columnwidth]
{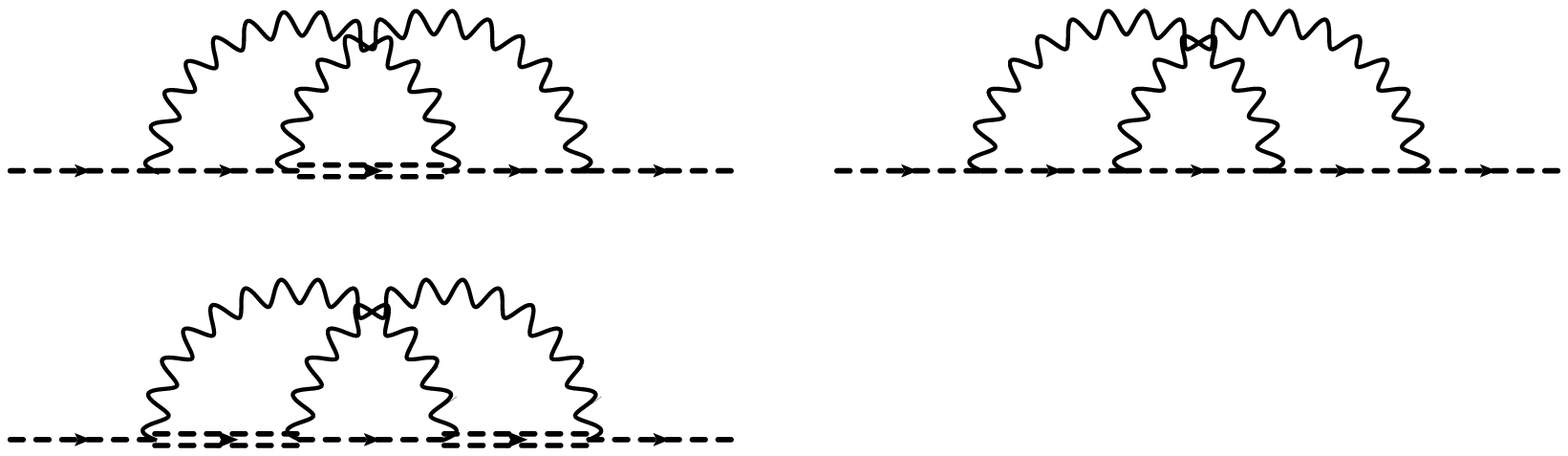}
\begin{picture}(0,0)
  \put(-235,55){\text{$\delta\Sigma_S (\epsilon) = $}}
  \put(-235,35){\text{$\delta\Sigma_T (\epsilon) = $}}  
  \put(-100,35){\text{$+$}}  
  \put(-100,55){\text{$+$}}    
 \put(-0,35){\text{$+$}}    
\end{picture}
\caption{Diagrams for the leading $1/\mathcal{N}$ corrections $\delta\Sigma_S(\epsilon)$ and $\delta\Sigma_T(\epsilon)$   (a) to singlet and (b) to triplet self-energies.  Double dashed line and dashed line represent two-fermion Green's function in singlet and triplet channels correspondingly.}\label{fig:delta_Sigma}
\end{figure}

Let us consider  $1/\mathcal{N}$ corrections to the self-energies $\Sigma_S(\epsilon)$ and $\Sigma_T(\epsilon)$, calculated in Self-Consistent Born Approximation (see Eqs. (\ref{eq:Bag_Physics_Dyson})). In order to do this we account for vertex corrections $\delta \Sigma_S(\epsilon)$ and $\delta \Sigma_T(\epsilon)$ to self-energies obtained in SCBA, corresponding diagrams are shown in Fig. \ref{fig:delta_Sigma}. 
The vertex correction to the singlet self-energy reads
\begin{equation}\label{eq:delta_Sigma_singlet}
\begin{split}
\delta \Sigma_S(\epsilon) = 3\sum_{\bm q, \bm k}|g_{ST}(q)|^2 |g_{ST}(k)|^2 G_{T,q}G_{T,k}G_{S,qk} - \\ 6\sum_{\bm q, \bm k}g_{ST}(q)g_{ST}^*(k)g_{TT}(k)g_{TT}^*(q) G_{T,q}G_{T,k}G_{T,qk}.
\end{split}
\end{equation}
The combinator factors come from contractions of the corresponding tensor structures of the effective vertices in Eq. (\ref{eq:Hamiltonian_singlet-triplet}).
In the similar way the vertex correction to the triplet self-energy is given by 
\begin{equation}\label{eq:delta_Sigma_triplet}
\begin{split}
&\delta\Sigma_T(\epsilon) = \sum_{\bm q, \bm k}|g_{ST}(q)|^2 |g_{ST}(k)|^2 G_{S,q} G_{S,k} G_{T,qk} - \\ &2\sum_{\bm q,\bm k}g_{TT}(q)g_{TT}^*(k) g_{ST}^*(q) g_{ST}(k) G_{T,q} G_{T,k} G_{S,qk} + \\  &2\sum_{\bm q, \bm k}|g_{TT}(q)|^2 |g_{TT}(k)|^2 G_{T,q} G_{T,k} G_{T,qk}.
\end{split}
\end{equation}

Here we are using shorten notations $G_{n, q} = G_n(\epsilon - \omega_q)$, $G_{n, k} = G_n(\epsilon - \omega_k)$ and $G_{n, qk} = G_n(\epsilon - \omega_q - \omega_k)$ for singlet and triplet Green's functions ($n=S,T$). One can check that in the limit $r\rightarrow \infty$ the correction $\delta\Sigma_S$ will be suppressed by the factor $1/\mathcal{N} = 1/3$ with respect to $\Sigma^{(2\,loop)}_{S,T}$ calculated in SCBA within two loop approximation.


The relative shift of binding energy, calculated with and without vertex correction, does not exceed $20\%$. It can be considered as a confirmation of applicability of $1/\mathcal{N}$ expansion for the effective "Lamb-shift" theory described by the Hamiltonian (\ref{eq:Hamiltonian_singlet-triplet}).

\end{document}